\documentclass[runningheads]{llncs}
\usepackage{graphicx}
\usepackage{subcaption}

\begin{document}
\title{Transferability limitations for Covid 3D Localization Using SARS-CoV-2 segmentation models in 4D CT images}
%
%

\author{Constantine Maganaris\inst{1}\orcidID{0000-0003-2738-684X} \and
Eftychios Protopapadakis\inst{1}\and
Nikolaos Bakalos\inst{2}\and
Nikolaos Doulamis\inst{1}\and
Dimitris Kalogeras\inst{2}\and
Aikaterini Angeli\inst{2}
}

\authorrunning{C. Maganaris et al.}
%
\institute{National Technical University of Athens, School of Rural, Surveying and Geoinformatics Engineering, Greece \and
Institute of Communications and Computer Systems, Athens, Greece}
\maketitle              
\begin{abstract}

In this paper, we investigate the transferability limitations when using deep learning models, for semantic segmentation of pneumonia-infected areas in CT images. The proposed approach adopts a 4 channel input; 3 channels based on Hounsfield scale, plus one channel (binary) denoting the lung area. We used 3 different, publicly available, CT datasets. If the lung area mask was not available, a deep learning model generates a proxy image. Experimental results suggest that transferability should be used carefully, when creating Covid segmentation models; retraining the model more than one times in large sets of data results in a decrease in segmentation accuracy.

\end{abstract}
\section{Introduction}
According to John Hopkins University (27 June 2022), the coronavirus pandemic (Covid) and its variations, has infected more than 543 million people and caused more than 6.3 million deaths around the globe \cite{noauthor_Covid-19_nodate}. Since the declaration of Public Health emergency of International Concern on January 30 2020 \cite{noauthor_statement_nodate} from the World Health Organization (WHO) there was a research outbreak in the field, to respond the emergency situation. A great amount of research has been focused since then on fast detection of Covid effects, using CT or X-rays of the thorax, in parallel with other methods like Antigenic and Reverse transcription-polymerase chain reaction (RT-PCR) testing that focus on the identification of the virus without any consideration of its effects.

Despite the progress on the medical image analysis field, it is crucial to continue developing automated decision making tools to assist healthcare personnel and overcome all the issues that comes with the analysis of Computed Tomography Imaging. One of the main setbacks in creating accurate models is the lack of publicly available datasets, with chest scans of infected people. Also, a lot of data that have been used so far, have faulty cleansing, so they create frustration on the results produced by deep learning models, trained to detect Covid \cite{noauthor_hundreds_nodate}. Moreover, scans coming from different equipment, produce data that differentiate on how they signal Covid areas using the Hounsfield scale thus making the training of appropriate models even more difficult due to the significant differences in the training data.

In our research we use three different datasets in order to detect the limitations that transferability of deep learning models has and estimate the accuracy of them to detect segments of Covid areas (i.e., markings of ground glass opacities, consolidation and pleural) \cite{cozzi_ground-glass_2021}. These datasets include 5,459 slides of CT scans with annotated Lung segments in two of the three sets, and masks with Covid signals to all the sets. 

To present our model predictions, we have visualized 3D reconstructions of the lungs using data from the CT scans. This way we achieved better understanding on how the model works in the complete CT scan and not by comparing separate slides of it. In our 3D model reproduction, we used only the lung areas of the slides, the annotations of the Covid areas made by the radiologists (ground truth) and the predicted Covid areas from our model. Now having these representations in hand we can view the bigger picture and how the model performs in a patient. This is making it easier for a user (i.e., a doctor) of the model to extract information from the results of how serious the illness is, or how the infection responds to medication etc. by comparing different scans of the same patience in different points of time. 


In order to extract our results and increase the number of data, we needed to annotate the lung areas of the third dataset, since it didn't include the lung masks in it. To achieve this, we trained a U-Net to identify the lung segments in CT scans of thorax. The model had quite good performance thus we used it to create the lung masks of the third dataset.

\section{Related work}
Deep learning methodologies using various types of images are common for identification, detection or segmentation in medical imaging \cite{voulodimos_deep_2018} and in biomedical applications \cite{le_fertility-gru_2019}. In this context, researchers already investigating several approaches to assist medical professionals with Covid detection.
An initial approach was to classify multiple CT slices using a convolutional neural network variation \cite{li_coronavirus_2020}. The adopted methodology is able to identify a viral infection with a Receiver Operating Characteristic/Area Under the Curve (ROCAUC) score of 0.95 (a score of 1 indicates a perfect classifier). However, despite the high detection rates, the authors indicated that it was extremely difficult to distinguish among different types of viral pneumonia based solely on CT analysis.

Convolutional Neural Network (CNN) variations for the distinction of coronavirus vs. non-coronavirus cases have been proposed by \cite{li_using_2020}. The specific approach allows for a distinction among Covid, other types of viral infections, and non-infection cases. Results indicate that there are adequate detection rates and a higher detection rate than RT-PCR testing. Towards this direction, CNN structures are combined with Long Short Term Memory (LSTM) networks to improve the classification accuracy of CNN networks further \cite{islam_combined_2020}. Additionally the work of \cite{fan_inf-net_2020} introduced a parallel partial decoder, called Inf-Net, which combines aggregation of high-level features to generate a global map. This is achieved through the use of convolutional hierarchies.

A U-Net-based model, named U-Net++, was applied to high-resolution CT images for Covid detection in \cite{chen_deep_2020}. Furthermore, in \cite{ardakani_application_2020} a system for the detection of Covid using 10 variants of CNNs in CT images is proposed, including AlexNet, VGG-16, VGG-19, SqueezeNet, GoogleNet, MobileNet-V2, ResNet-18, ResNet-50, ResNet-101, and Xception. ResNet-101 and Xception outperformed the remaining ones. AlexNet and Inception-V4 were also used for Covid detection in CT scans in \cite{cifci_deep_2020}. The framework presented in \cite{singh_classification_2020} used a CNN and an Artificial Neural Network Fuzzy Inference System (ANNFIS) to detect Covid.
Focusing on segmentation, a type 2 fuzzy clustering system combined with a Super-pixel based Fuzzy Modified Flower Pollination Algorithm is proposed in \cite{chakraborty_sufmofpa_2021} for Covid CT image segmentation.

In \cite{katsamenis_transfer_2020} the experimental results indicate that the transfer learning approach outperforms the performance obtained without transfer learning, for the Covid classification task in chest X-ray images, using deep classification models, such as convolutional neural networks (CNNs).

Finally in our previous research \cite{10.1145/3529190.3534736}, we used the first two of the three datasets we are using now, to train a U-Net using only 2 channels of input in each image, the first channel included the normalized Hounsfield values of the CT and the second channel had the lung mask of the specific CT slide. Compared to our previous results we found that there was a significant drop in performance when we retrained our model using an additional dataset. 

\subsection{Our contribution}
Compared to the work of \cite{10.1145/3529190.3534736} in evaluating transferability of deep learning models in Covid segmentation to Computed Tomography data, we introduce a new architecture concept and we run additional evaluation tests. At first, our approach uses 4-channel inputs; 3 Hounsfield based channels see subsec. \ref{DatasetDescription} plus a binary one, indicating the lung area. Secondly, we run an extensive evaluation on more datasets to investigate when transferability is not beneficial. In this scenario, we considered a successive retraining process and we investigate the performance level deterioration, over the original dataset, once retraining the model is completed. 
Experimental results, sec. \ref{ExperimentalResults}, indicate that transferability should be used with cautious, since it is not always beneficial.   

\section{Experimental setup}

\begin{figure}[hbt!]
     \centering
         \centering
         \includegraphics[width=\textwidth]{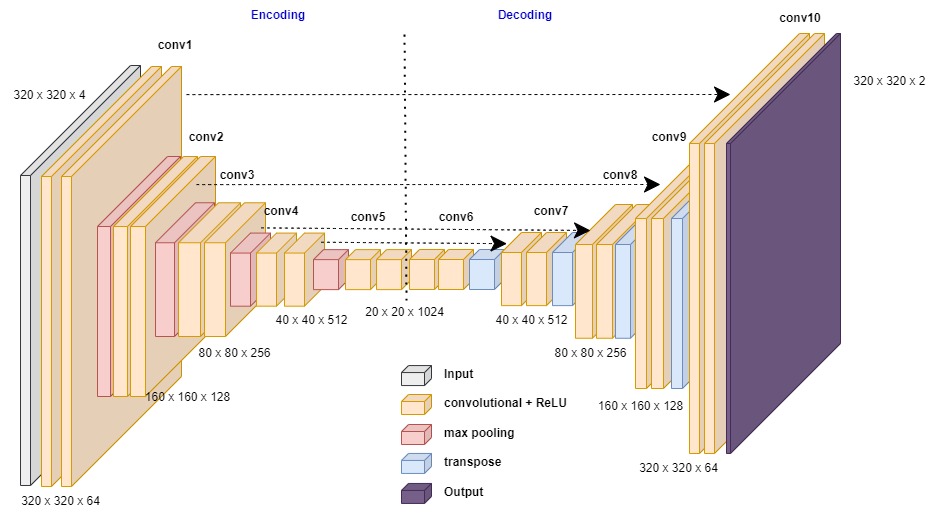}
         \caption{U-Net Architecture}
         \label{fig:U-Net Architecture}
\end{figure}

The U-Net and the data transformation scripts were developed in Python 3 using TensorFlow and Keras libraries. The models were trained in a VM using Unix MATE OS with 8 Core CPU and 64GB RAM provided GRNet Synefo service. Figure \ref{fig:U-Net Architecture} presents the architecture of the U-Net model used.

\subsection{Datasets description}
\label{DatasetDescription}
To extract our results, we used three Lung Covid infected datasets, the Covid CT segmentation dataset \cite{noauthor_Covid-19_nodate-1} from which we used only the Segmentation dataset nr. 2, this includes 9 DICOM files of continuous lung CT scans and the 20th April update \cite{jun_Covid-19_2020}  which contains another 20 labeled Covid CT scans, from this we used only the 10 files marked as Coronacases and not the Radeopaedia ones. 
The reason for this is that all DICOM files we used contained data in the Hounsfield scale \cite{PMID:31613501} the Radeopaedia set of DICOM files contained pixel values in the range of 0-255, therefore we could not use it since it did not follow our normalization procedure.
The first set of 9 DICOM files (we refer to this set as CT 1-9) contained 829 slides of CT, having dimensions of 630x630 pixels and contained already hand annotated lung and Covid masks for each slide of it. Similarly, the Coronacases dataset contained 10 CT scans (we refer to this set as CT 10) with 2581 slides in total having dimensions of 512x512 and included also annotated masks by radiologists of the lung and Covid areas. Both sets, include continuous slides of complete lung CT scans of the same patient and not slides of different patients in each DICOM file. 
The third dataset we used was the MosMedData  \cite{Morozov2020.05.20.20100362} which includes 254 CT scans of healthy lungs and 50 scans infected with Covid. From this set we used only the 50 scans infected with Covid (we refer to this set as CT-1), since there is a large number of slides of healthy areas even in this set (963 slides). The slides were having dimensions of 512x512 pixels and have hand annotated Covid masks but no lung masks, therefore we predicted the lung masks using a U-Net we created for this specific purpose.

To construct our final dataset we used only the slides that include lung areas, in order to achieve better results by reducing the extra information of slides without lung areas. Please note that for CT 1-9 dataset we have 714 masks with Covid (214 train + 60 validation + 440 test slides), but only 713 masks with lung. This is attributed to a human annotation error, see fig. \ref{fig:False Covid annotations}, resulting in one additional image.

The splitting of the dataset to train/val/test set did not consider leaving a test subject (entire patient's CT) out for test.

\begin{table}[!h]
\small
\caption{Datasets' properties}
\centering
\resizebox{\textwidth}{!}{%
\begin{tabular}{ccccccccc}
\hline
\multicolumn{1}{c}{\textbf{Set Name}} &
  \multicolumn{1}{c}{\textbf{\begin{tabular}[c]{@{}c@{}}\# DICOM Files\end{tabular}}} &
  \multicolumn{1}{c}{\textbf{\# Slides}} &
  \multicolumn{1}{c}{\textbf{\begin{tabular}[c]{@{}c@{}}Slides with\\ lung areas\end{tabular}}} &
  \multicolumn{1}{c}{\textbf{\begin{tabular}[c]{@{}c@{}}Slides with \\ Covid areas\end{tabular}}} &
  \multicolumn{1}{c}{\textbf{\begin{tabular}[c]{@{}c@{}}Train set slides\\ (ratio 50\%/50\% \\ Covid/non-Covid slides)\end{tabular}}} &
  \multicolumn{1}{c}{\textbf{\begin{tabular}[c]{@{}c@{}}Validation set slides \\ (ratio 50\%/50\% \\ Covid/non-Covid slides)\end{tabular}}} &
  \multicolumn{1}{c}{\textbf{\begin{tabular}[c]{@{}c@{}}Test set \\Slides\end{tabular}}} &
  \multicolumn{1}{c}{\textbf{Epochs}} \\ \hline
CT 1-9 &
  9 &
  829 &
  713 &
  373 &
  440 &
  60 &
  214 &
  154 \\ \hline
CT 10 &
  10 &
  2581 &
  2156 &
  1351 &
  440 &
  60 &
  1656 &
  41 \\ \hline
  CT-1 &
  50 &
  2049 &
  1748 &
  785 &
  440 &
  60 &
  1248 &
  39
  \\ \hline
  \label{table:1}
\end{tabular}}
\end{table}

For the normalization process, we resized the images to 320x320 pixels using Nearest Neighbor Interpolation and we kept only the Hounsfield values in the range of -970 to -150. We reformatted the CT data to 3 channels, using different intervals, Channel 1 was [-970,-150], Channel 2 was [-700,-450] and Channel 3 was [-450,-150]. These intervals came from empirical analysis of the CT images, knowing that healthy lung areas are close to [-700,-600] \cite[pg. 379]{Kazerooni2003-tj} and checking also that ground glass opacities, consolidation and pleura are covered in the areas of [-970,-150].

We tried also a different approach, to break the intervals to [-970,-700], [-700,-450] and [-450,-150] but this did not produced output results that could be used with any decimal threshold to create binary images of Covid masks. We overcome this issue by creating the overlapping channel of the intervals. In the approach we use now the 1\textsuperscript{st} channel overlaps with the other 2 channels of the CT data. We normalized each pixel based on its value, using the following types for the first (\ref{eq:1}), the second (\ref{eq:2}) and the third (\ref{eq:3}) channel:

\begin{equation} \label{eq:1}
\frac{(pixelVal+970)}{(-150+970)}
\end{equation}

\begin{equation} \label{eq:2}
\frac{(pixelVal+700)}{(-450+700)}
\end{equation}

\begin{equation} \label{eq:3}
\frac{(pixelVal+450)}{(-150+450)}
\end{equation}

For every normalized pixel value we get that is greater than 1, we assign the value 1 and every value less than 0, we assign the value 0. This way our dataset includes values only in the range [0,1] in all three channels. 
The radiologists have also marked in separate files the lung masks and the Covid masks for each slide of the CT scans we used. Therefore we arranged our Training input in the form of $n \times 320 \times 320 \times 4$ since in the first 3 channels use the normalized values of the CT slides in the ranges beforementioned and in the 4\textsuperscript{th} channel we used the lung masks of each slide in binary values of 0,1 (1 in pixels that are marked as lung, 0 elsewhere). The output was having the form of $n \times 320 \times 320 \times 2$ signaled with binary values 0,1. In the first channel we annotate using 1 in pixels that are marked as Covid, 0 elsewhere and in the second channel we mark 1 the non-Covid areas and 0 elsewhere.

\subsection{Experimental results}
\label{ExperimentalResults}
With the pre-described formation of all datasets, we started the training of a U-Net using the CT 1-9 data, with 4 encoding/decoding layers having as input $440 \times 320 \times 320 \times 4$. In our input data, we have in the 1\textsuperscript{st}-3\textsuperscript{rd} channels the data of the CT scan using the channels described above in normalized values and in the 4\textsuperscript{th} channel we have the lung mask of the specific CT slide. This way the model focuses only in the lung areas of the CT scans in the learning process. In the output we have $440 \times 320 \times 320 \times 2$ channels. In the first channel we mark the slide using [0,1] and in the second one we have the inverted [1,0] masks data of the Covid areas. We also used a validation set of $60 \times 320 \times 320 \times 4$ as input and $60 \times 320 \times 320 \times 2$ as output. In the U-Net (i.e., Figure \ref{fig:U-Net Architecture}) we used the rectified linear activation unit (ReLU) function for the 3x3 conv layers and the Sigmoid activation in the 1x1 conv layer, to get output values in the range (0,1), a learning rate of 0.0001, a batch size of 45 and the shuffling enabled. The max epochs were 200 from which it only used 154 until the early stopping engaged. 

In order to objectively evaluate our results, four different metrics are considered: accuracy, precision, recall, and the F1‐score. We extracted these metrics for the test data of the CT 1-9 set (i.e., 214 slides) (see Table \ref{table:PerfMetrics}), from which we got an Accuracy value with an average of 0.9960 and an F1 score with an average of 0.7993 only using CT slides that include Covid areas. 

\begin{table}[h!]
\center
\caption{Performance Metrics (Tr: Training, R: Retrain dataset, Te: Test dataset)}
\resizebox{0.8\textwidth}{!}{%
\begin{tabular}{p{3cm}lcccc}
\cline{2-5}
\multicolumn{1}{l}{} & Acc & Pre & Rec   & F1       \\ \hline
Tr1\_R\_None\_Te1  & 0.996067	& 0.825346 & 0.801561 & 0.799373\\
Tr1\_R\_None\_Te2 & 0.990696	& 0.627217	& 0.447298 &	0.477114 \\
Tr1\_R\_None\_Te3 & 0.995938 &	0.605107	& 0.408050	& 0.435020 \\\hline
Tr1\_R\_2\_Te2 & 0.996483 &	0.811026 &	0.801767 &	0.791740 \\
Tr1\_R\_2\_Te1 & 0.991956 &	0.725649 &	0.714044 &	0.690094 \\
Tr1\_R\_2\_Te3 & 0.996360	& 0.619701 &	0.435246 &	0.454912 \\\hline
Tr1\_R\_2\_R\_3\_Te3 & 0.996989 &	0.684802 &	0.478032 &	0.514396 \\
Tr1\_R\_2\_R\_3\_Te1 & 0.986916	& 0.404010 &	0.209785 &	0.242582 \\
Tr1\_R\_2\_R\_3\_Te2 & 0.992636	& 0.817556	& 0.492780	& 0.572272 \\\hline
Tr2\_R\_None\_Te2	& 0.993587 &	0.611667 &	0.887216 &	0.705930 \\
Tr2\_R\_2\_Te3 & 0.994351	& 0.453238 & 0.817842 &	0.542704 \\
Tr2\_R\_2\_Te2 & 0.992053 & 0.598488 &	0.828759 & 0.651196\\\hline
Tr3\_R\_None\_Te3 &	0.996315 &	0.566403 &	0.732628 &	0.598084 
\label{table:PerfMetrics}
\end{tabular}}
\end{table}

Afterwards we moved with the extraction of the same metrics, using the test slides of the second dataset (CT 10). From these we got an average accuracy of 0.9907  but an F1 score of 0.4771. Then we re-trained our model using a small portion of this dataset. From CT 10 dataset we had a similar size of train and validation sets (440 and 60) and we continued the train of the previous model (2\textsuperscript{nd} version) but now we reduced the max epochs to 50. From these epochs the train process used 41 until the early stopping engaged. We used the re-trained model in the test set of the 2\textsuperscript{nd} dataset slides to obtain our metrics.

From the extraction of metrics, the retrained model returned an average accuracy of 0.9965 and an average F1 score of 0.7917 which was an improvement compared to the 0.4771 F1 before the retrain of the model. We then used the retrained model to the test set of the first dataset and got an average accuracy of 0.99196 and an average F1 score of 0.6901. Here we note a drop of 0.1 in F1 score compared with the score of the first model in this dataset. 

As a next step we extracted an average accuracy of 0.9964, using the second model to the 3\textsuperscript{rd} dataset and an average F1 of 0.4549. Then we retained the 2\textsuperscript{nd} model using the 3\textsuperscript{rd} dataset for 39 epocs until early stopping engaged. Then we used the 3\textsuperscript{rd} model to extract an average accuracy of 0.9970 and an F1 average of 0.5144 from the 3\textsuperscript{rd} dataset. We, also, used the 3\textsuperscript{rd} model to extract metrics from the two first test sets. For the first test set, we got an average accuracy of 0.9870 and an average F1 of 0.2426 and for the second test set we got an average accuracy of 0.9926 and an average F1 of 0.5723.

\begin{figure}[!h]
     \centering
     \begin{subfigure}[b]{0.3\columnwidth}
         \centering
         \includegraphics[width=25mm,height=25mm]{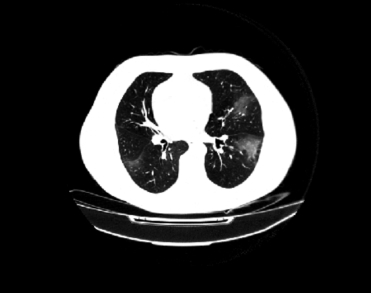}
     \end{subfigure}
     \hfill
     \begin{subfigure}[b]{0.3\columnwidth}
         \centering
         \includegraphics[width=25mm,height=25mm]{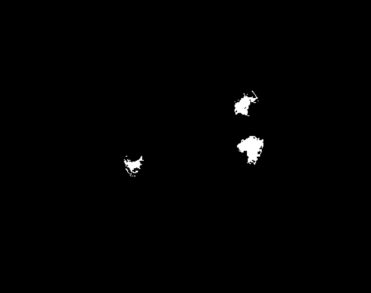}
     \end{subfigure}
     \hfill
     \begin{subfigure}[b]{0.3\columnwidth}
         \centering
         \includegraphics[width=25mm,height=25mm]{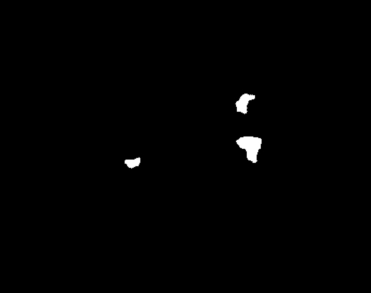}
     \end{subfigure}
        \caption{3\textsuperscript{rd} Dataset Good Prediction Example; CT scan (left), Ground Truth (middle), Prediction (right)}
        \label{fig:3rd Dataset Good Prediction Example}
\end{figure}

The models' names notion was build to demonstrate the experiment process. A typical name consists of 3 parts: the dataset used for the training denoted with (Tr), the dataset(s) used for retraining, denoted with (R) and, lastly the test set, denoted with (Te). As such, a model name Tr1\_R\_2\_Te1 indicates a model initially trained on dataset 1, then used to initialize a second model, (re)trained on dataset 2, and finally got tested on the test set of dataset 1.
In Figure \ref{fig:3rd Dataset Good Prediction Example} we can see that the model performs great in the test set of the 3\textsuperscript{rd} dataset in Tr1\_R\_2\_R\_3\_Te3,  in a good scenario where the F1 Score of the image is 0.8456 for this specific image. 

\begin{figure}[hbt!]
     \centering
     \begin{subfigure}[b]{0.3\columnwidth}
         \centering
         \includegraphics[width=25mm,height=25mm]{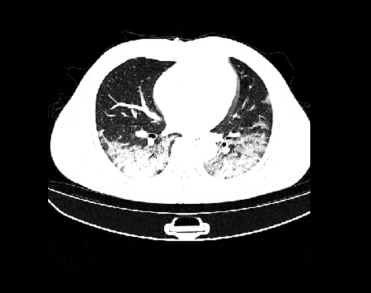}
     \end{subfigure}
     \hfill
     \begin{subfigure}[b]{0.3\columnwidth}
         \centering
         \includegraphics[width=25mm,height=25mm]{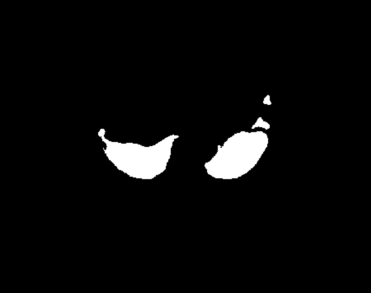}
     \end{subfigure}
     \hfill
     \begin{subfigure}[b]{0.3\columnwidth}
         \centering
         \includegraphics[width=25mm,height=25mm]{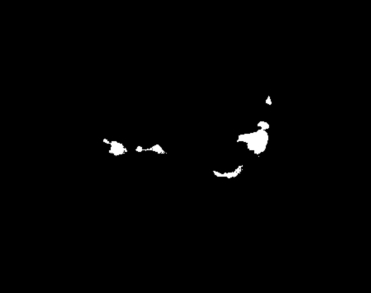}
     \end{subfigure}
        \caption{2\textsuperscript{nd} Dataset Bad Prediction Example; CT scan (left), Ground Truth (middle), Prediction (right)}
        \label{fig:2nd Dataset Bad Prediction Example}
\end{figure}

Results indicate that when forcing a retrain to a new data set, the existing model achieves an increase in performance scores (e.g. F1 score, for dataset CT-1, from 0.45 to 0.51), at the cost of a major decrease on the initial datasets' performance scores (e.g. F1 score, for dataset CT 1-9, from 0.69 to 0.24 and for dataset CT 10, from 0.79 to 0.57). 
In Figure \ref{fig:2nd Dataset Bad Prediction Example} we see an example of how the re-trained model Tr1\_R\_2\_R\_3\_Te2 performs, in a bad scenario where the F1 Score of the image is 0.3571.

We also trained a U-Net (i.e., Tr2\_R\_None\_Te2) using the 2\textsuperscript{nd} dataset (i.e., CT 10) as input. With this model on the test data we got an average accuracy of 0.9936 and an F1 of 0.7059. We retrained this model using the train data of dataset three (CT-1) and we got an average accuracy of 0.9944 and 0.5427 F1 score. With this model we extracted again the metrics for the 2\textsuperscript{nd} set test data (i.e., Tr3\_R\_2\_Te\_2) and it returned an average accuracy of 0.9921 and 0.6512 F1.

Furthermore we trained a new U-Net (i.e., Tr3\_R\_None\_Te3) having the same architecture as the previous one, using the train/validation set of the 3\textsuperscript{rd} dataset (i.e., CT-1). We then extracted the performance metrics of the CT-1 test data, and we got an average of accuracy 0.9963 and 0.5981 F1 score.
 
Table \ref{table:PerfMetrics} indicates the difference in performance when successive retraining process took place. We observe a significant drop in performance when an already retrained model is transferred for test on a new dataset (i.e., Tr1\_R\_2\_R\_3\_Te1 and Tr1\_R\_2\_R\_3\_Te2). It appears that mixing multiple datasets deteriorates the performance on the already established ones, with minor or no gain on the new dataset.

We must note here that all the slides that have been used from all datasets, had an area marked as lung in the accompanied mask files from the radiologists. If a slide did not include a lung area, it was not included in the any of the trained/validation or test set data. Also all the average metrics were gathered per CT slide, and concern only slides with marked Covid areas.

\section{Dataset limitations}
\begin{figure}[!hbt]
     \centering
     \begin{subfigure}[b]{0.40\textwidth}
         \centering
         \includegraphics[width=\textwidth]{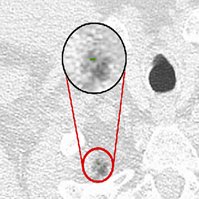}
         \label{fig: False annotated area 1}
     \end{subfigure}
     \hfill
     \begin{subfigure}[b]{0.40\textwidth}
         \centering
         \includegraphics[width=\textwidth]{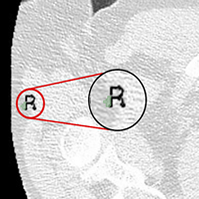}
         \label{fig:False annotated area 2}
     \end{subfigure}
       \caption{False Covid annotations; Area 1 (left), Area 2 (right)}
        \label{fig:False Covid annotations}
\end{figure}

In the datasets we have used, we spotted some inconsistencies, since its annotation was made by hand. Specifically in our pre-train analysis, we found that in DICOM files there were slides with a false marked Covid areas. As we can see in Figure \ref{fig:False Covid annotations} the radiologist has a marked Covid area of 3 pixels by mistake (Area 1), or in a section that there is no lung at all (Area 2).

Taking into account the previously mentioned paradigm, we cannot be sure of the extent of these faulty annotated areas, since the only chance to find them is to re-evaluate all annotations by a different radiologist and compare the results.

\section{3D representations}
To assist the medical personnel which works with tools like Computed Tomography scans, we constructed a Python script that exports a 3D representation of the CT scan data and the Covid segments produced by our model. 
At first, we extracted for every slide in a CT the color value of each pixel. We created an array of color values only in the lung areas, each entry of the array was having the form of (x position, y position, z position, color value). The z position was computed starting from 0 and adding a constant value for every new slide that we export its data. We did the same procedure for the Covid masks that our model predicted and for the ground truth, but in these the color value was in binary (i.e., 0 or 1). We then exported these arrays in comma separated files which we then imported to the ParaView visualizer \cite{noauthor_paraview_nodate} using the table to point filtering. 

\begin{figure}[hbt!]
     \centering
     \begin{subfigure}{0.3\textwidth}
         \centering
         \includegraphics[width=\textwidth]{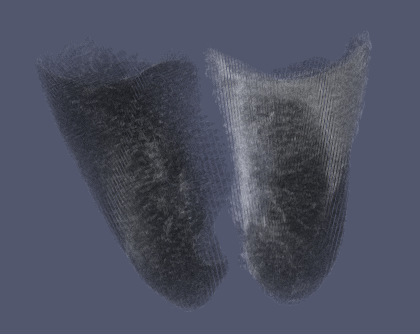}
     \end{subfigure}
     \hfill
     \begin{subfigure}{0.3\textwidth}
         \centering
         \includegraphics[width=\textwidth]{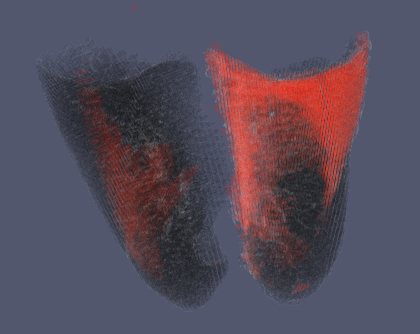}
     \end{subfigure}
     \hfill
     \begin{subfigure}{0.3\textwidth}
         \centering
         \includegraphics[width=\textwidth]{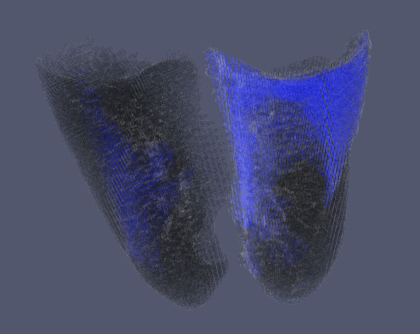}
     \end{subfigure}
        \caption{3D Representation; CT scan (left), Ground Truth (middle), Prediction (right)}
        \label{fig:3D Representation}
\end{figure}

With the 3D reconstruction, the evaluation of a patience status is easier for the medical personnel, as they can see the whole lung and not evaluate separate CT slides which can be significant in number. With this tool it is also easier to estimate how a treatment performs in time, using different 3D representations of the same patient.

\section{Conclusions}
In this paper we present the transferability of deep learning models for semantic segmentation of pneumonia infected areas in CT images. The limitations of such transferability was studied, when retraining the model multiple times in a portion of data of multiple sets for a short number of epochs. It is important to note that there is a significant drop in prediction outcome and with metrics (i.e., precision, recall and F1) when the model is trained in more than two sets. The main reason for this is that the CT scans data come from different sources thus when the model is retrained in new sets of data, it may increase its performance for the current set, but worsen its performance in previous ones. 
However, the performance of our model still showcases high performance metrics, even in instances where there are multiple retrainings over the initial model. However, in these cases the selection of the initial dataset greatly affects the performance. This is due to the lack of appropriate annotation in a lot of datasets, which makes them unsuitable for the initial development of the model.

\bibliographystyle{splncs04}
\bibliography{bibliography}

\begin{thebibliography}{10}
\providecommand{\url}[1]{\texttt{#1}}
\providecommand{\urlprefix}{URL }
\providecommand{\doi}[1]{https://doi.org/#1}

\bibitem{noauthor_Covid-19_nodate}
{COVID}-19 {Map} (2022), \url{https://coronavirus.jhu.edu/map.html}

\bibitem{noauthor_statement_nodate}
Statement on the tenth meeting of the {International} {Health} {Regulations}
  (2005) {Emergency} {Committee} regarding the coronavirus disease ({COVID}-19)
  pandemic (2022),
  \url{https://www.who.int/news/item/19-01-2022-statement-on-the-tenth-meeting-of-the-international-health-regulations-(2005)-emergency-committee-regarding-the-coronavirus-disease-(covid-19)-pandemic}

\bibitem{noauthor_hundreds_nodate}
Hundreds of {AI} tools have been built to catch covid. {None} of them helped.
  (2022),
  \url{https://www.technologyreview.com/2021/07/30/1030329/machine-learning-ai-failed-covid-hospital-diagnosis-pandemic/}

\bibitem{ardakani_application_2020}
Ardakani, A.A., Kanafi, A.R., Acharya, U.R., Khadem, N., Mohammadi, A.:
  Application of deep learning technique to manage {COVID}-19 in routine
  clinical practice using {CT} images: {Results} of 10 convolutional neural
  networks. Computers in Biology and Medicine  \textbf{121},  103795 (Jun
  2020). \doi{10.1016/j.compbiomed.2020.103795},
  \url{https://www.sciencedirect.com/science/article/pii/S0010482520301645}

\bibitem{chakraborty_sufmofpa_2021}
Chakraborty, S., Mali, K.: {SuFMoFPA}: {A} superpixel and meta-heuristic based
  fuzzy image segmentation approach to explicate {COVID}-19 radiological
  images. Expert Systems with Applications  \textbf{167},  114142 (Apr 2021).
  \doi{10.1016/j.eswa.2020.114142},
  \url{https://www.sciencedirect.com/science/article/pii/S0957417420308897}

\bibitem{chen_deep_2020}
Chen, J., Wu, L., Zhang, J., Zhang, L., Gong, D., Zhao, Y., Chen, Q., Huang,
  S., Yang, M., Yang, X., Hu, S., Wang, Y., Hu, X., Zheng, B., Zhang, K., Wu,
  H., Dong, Z., Xu, Y., Zhu, Y., Chen, X., Zhang, M., Yu, L., Cheng, F., Yu,
  H.: Deep learning-based model for detecting 2019 novel coronavirus pneumonia
  on high-resolution computed tomography. Scientific Reports  \textbf{10}(1),
  19196 (Nov 2020). \doi{10.1038/s41598-020-76282-0},
  \url{https://www.nature.com/articles/s41598-020-76282-0}, number: 1
  Publisher: Nature Publishing Group

\bibitem{cifci_deep_2020}
Cifci, M.: Deep {Learning} {Model} for {Diagnosis} of {Corona} {Virus}
  {Disease} from {CT} {Images}. International Journal of Scientific Research
  and Management  \textbf{11}, ~273 (Apr 2020)

\bibitem{cozzi_ground-glass_2021}
Cozzi, D., Cavigli, E., Moroni, C., Smorchkova, O., Zantonelli, G., Pradella,
  S., Miele, V.: Ground-glass opacity ({GGO}): a review of the differential
  diagnosis in the era of {COVID}-19. Japanese Journal of Radiology
  \textbf{39}(8),  721--732 (2021). \doi{10.1007/s11604-021-01120-w},
  \url{https://www.ncbi.nlm.nih.gov/pmc/articles/PMC8071755/}

\bibitem{PMID:31613501}
DenOtter, T.D., Schubert, J.: Hounsfield Unit. StatPearls Publishing, Treasure
  Island (FL) (2021), \url{http://europepmc.org/books/NBK547721}

\bibitem{fan_inf-net_2020}
Fan, D.P., Zhou, T., Ji, G.P., Zhou, Y., Chen, G., Fu, H., Shen, J., Shao, L.:
  Inf-{Net}: {Automatic} {COVID}-19 {Lung} {Infection} {Segmentation} {From}
  {CT} {Images}. IEEE Transactions on Medical Imaging  \textbf{39}(8),
  2626--2637 (Aug 2020). \doi{10.1109/TMI.2020.2996645}, conference Name: IEEE
  Transactions on Medical Imaging

\bibitem{islam_combined_2020}
Islam, M.Z., Islam, M.M., Asraf, A.: A combined deep {CNN}-{LSTM} network for
  the detection of novel coronavirus ({COVID}-19) using {X}-ray images.
  Informatics in Medicine Unlocked  \textbf{20},  100412 (Jan 2020).
  \doi{10.1016/j.imu.2020.100412},
  \url{https://www.sciencedirect.com/science/article/pii/S2352914820305621}

\bibitem{jun_Covid-19_2020}
Jun, M., Cheng, G., Yixin, W., Xingle, A., Jiantao, G., Ziqi, Y., Minqing, Z.,
  Xin, L., Xueyuan, D., Shucheng, C., Hao, W., Sen, M., Xiaoyu, Y., Ziwei, N.,
  Chen, L., Lu, T., Yuntao, Z., Qiongjie, Z., Guoqiang, D., Jian, H.:
  {COVID}-19 {CT} {Lung} and {Infection} {Segmentation} {Dataset} (Apr 2020).
  \doi{10.5281/zenodo.3757476}, \url{https://zenodo.org/record/3757476}, type:
  dataset

\bibitem{katsamenis_transfer_2020}
Katsamenis, I., Protopapadakis, E., Voulodimos, A., Doulamis, A., Doulamis, N.:
  Transfer {Learning} for {COVID}-19 {Pneumonia} {Detection} and
  {Classification} in {Chest} {X}-ray {Images}. Tech. rep., medRxiv (Dec 2020).
  \doi{10.1101/2020.12.14.20248158},
  \url{https://www.medrxiv.org/content/10.1101/2020.12.14.20248158v1}, type:
  article

\bibitem{Kazerooni2003-tj}
Kazerooni, E.A., Gross, B.H.: The core curriculum. Core Curriculum Series,
  Lippincott Williams and Wilkins, Philadelphia, PA (Sep 2003), pg. 379

\bibitem{le_fertility-gru_2019}
Le, N.Q.K.: Fertility-{GRU}: {Identifying} {Fertility}-{Related} {Proteins} by
  {Incorporating} {Deep}-{Gated} {Recurrent} {Units} and {Original}
  {Position}-{Specific} {Scoring} {Matrix} {Profiles}. Journal of Proteome
  Research  \textbf{18}(9),  3503--3511 (Sep 2019).
  \doi{10.1021/acs.jproteome.9b00411},
  \url{https://doi.org/10.1021/acs.jproteome.9b00411}, publisher: American
  Chemical Society

\bibitem{li_using_2020}
Li, L., Qin, L., Xu, Z., Yin, Y., Wang, X., Kong, B., Bai, J., Lu, Y., Fang,
  Z., Song, Q., Cao, K., Liu, D., Wang, G., Xu, Q., Fang, X., Zhang, S., Xia,
  J., Xia, J.: Using {Artificial} {Intelligence} to {Detect} {COVID}-19 and
  {Community}-acquired {Pneumonia} {Based} on {Pulmonary} {CT}: {Evaluation} of
  the {Diagnostic} {Accuracy}. Radiology  \textbf{296}(2),  E65--E71 (Aug
  2020). \doi{10.1148/radiol.2020200905},
  \url{https://pubs.rsna.org/doi/10.1148/radiol.2020200905}, publisher:
  Radiological Society of North America

\bibitem{li_coronavirus_2020}
Li, Y., Xia, L.: Coronavirus {Disease} 2019 ({COVID}-19): {Role} of {Chest}
  {CT} in {Diagnosis} and {Management}. American Journal of Roentgenology
  \textbf{214}(6),  1280--1286 (Jun 2020). \doi{10.2214/AJR.20.22954},
  \url{https://www.ajronline.org/doi/10.2214/AJR.20.22954}, publisher: American
  Roentgen Ray Society

\bibitem{10.1145/3529190.3534736}
Maganaris, C., Protopapadakis, E., Bakalos, N., Doulamis, N., Kalogeras, D.,
  Angeli, A.: Evaluating transferability for covid 3d localization using ct
  sars-cov-2 segmentation models. In: Proceedings of the 15th International
  Conference on PErvasive Technologies Related to Assistive Environments. p.
  615–621. PETRA '22, Association for Computing Machinery, New York, NY, USA
  (2022). \doi{10.1145/3529190.3534736},
  \url{https://doi.org/10.1145/3529190.3534736}

\bibitem{noauthor_Covid-19_nodate-1}
{COVID}-19 (2022), \url{http://medicalsegmentation.com/covid19/}

\bibitem{Morozov2020.05.20.20100362}
Morozov, S., Andreychenko, A., Pavlov, N., Vladzymyrskyy, A., Ledikhova, N.,
  Gombolevskiy, V., Blokhin, I., Gelezhe, P., Gonchar, A., Chernina, V.:
  Mosmeddata: Chest ct scans with covid-19 related findings dataset. medRxiv
  (2020). \doi{10.1101/2020.05.20.20100362},
  \url{https://www.medrxiv.org/content/early/2020/05/22/2020.05.20.20100362}

\bibitem{singh_classification_2020}
Singh, D., Kumar, V., {Vaishali}, Kaur, M.: Classification of {COVID}-19
  patients from chest {CT} images using multi-objective differential
  evolution–based convolutional neural networks. European Journal of Clinical
  Microbiology \& Infectious Diseases  \textbf{39}(7),  1379--1389 (Jul 2020).
  \doi{10.1007/s10096-020-03901-z},
  \url{https://doi.org/10.1007/s10096-020-03901-z}

\bibitem{voulodimos_deep_2018}
Voulodimos, A., Doulamis, N., Doulamis, A., Protopapadakis, E.: Deep {Learning}
  for {Computer} {Vision}: {A} {Brief} {Review}. Computational Intelligence and
  Neuroscience  \textbf{2018},  e7068349 (Feb 2018).
  \doi{10.1155/2018/7068349},
  \url{https://www.hindawi.com/journals/cin/2018/7068349/}, publisher: Hindawi

\bibitem{noauthor_paraview_nodate}
{ParaView} (2022), \url{https://www.paraview.org/}

\end{thebibliography}

\end{document}